\documentclass[XeLaTex,twocolumn,epjc3]{svjour3}          

\RequirePackage[T1]{fontenc}

\smartqed  

\RequirePackage{graphicx}
\RequirePackage{mathptmx}      
\RequirePackage{flushend}
\usepackage{epstopdf}
\RequirePackage[numbers,sort&compress]{natbib}
\RequirePackage[colorlinks,citecolor=blue,urlcolor=blue,linkcolor=blue]{hyperref}

\journalname{Eur. Phys. J. C}

\begin{document}

\title{A family of well-behaved Karmarkar spacetime describing interior of relativistic stars}

\author{Ksh. Newton Singh\thanksref{e1,addr1}
        \and
        Neeraj Pant\thanksref{e2,addr2} 
        }

\thankstext{e1}{e-mail: ntnphy@gmail.com}
\thankstext{e2}{e-mail: neeraj.pant@yahoo.com}

\institute{Department of Physics, National Defence Academy, Khadakwasla, Pune-411023, India.\label{addr1}
 \and
        Department of Mathematics, National Defence Academy, Khadakwasla, Pune-411023, India.\label{addr2}}

\date{Received: date / Accepted: date}

\maketitle

\begin{abstract}
We are presenting a family of new exact solutions for relativistic anisotropic stellar objects by considering four dimensional spacetime embedded in five dimensional Pseudo Euclidean space known as Class I solutions. These solutions are well-behaved in all respects, satisfy all energy conditions and the resulting compactness parameter is also within Buchdahl limit. The well-behaved nature of the solutions for a particular star solely depends on index $n$. We have discussed the solutions in detail  for the neutron star XTE J1739-285 ($M=1.51M_\odot, ~R=10.9$ km). For this particular star, the solution is well behaved in all respects for $8 \le n \le 20$. However, the solutions with $n<8$ possess increasing trend of sound speed and the solutions belong to $n>20$ disobey causality condition. Further, the well-behaved nature of the solutions for PSR J0348+0432 ($2.01M_\odot, ~11$ km), EXO 1785-248 (1.3$M_\odot$, 8.85 km) and Her X-1 (0.85$M_\odot$, 8.1 km) are specified by the index $n$ with limits  $24 \le n \le 54$,  $1.5 \le n \le 4$ and $0.8 \le n \le 2.7$ respectively.
\end{abstract}

\section{Introduction}
The century-old search for exact solutions of the Einstein field equations began with Karl Schwarzschild obtaining vacuum solution describing the exterior of a spherically symmetric matter distribution \cite{karl1}. A natural line of pursuit would be to find an interior solution which matched smoothly to the Schwarzschild exterior solution. This internal solution was obtained by Schwarzschild in which assumed that the internal matter content of a spherical mass distribution was characterized by uniform density \cite{karl2}. Observations of stars and the understanding of particle physics within dense cores necessitated the search for more realistic solutions of the field equations. The inclusion of pressure anisotropy, charge, bulk viscosity, an equation of state, multilayered fluids and the departure from spherical symmetry has led to the discovery of hundreds of exact solutions describing relativistic stars in the static limit \cite{bowers,beken,sunil1,tik1,escu}. With the discovery of the Vaidya solution, it became necessary to model the gravitational collapse of radiating stars \cite{vaidya}. Since the star is dissipating energy in the form of a radial heat flux, the pressure at the boundary of the star is proportional to the outgoing heat flux as opposed to vanishing surface pressure in the non-dissipative case \cite{santos}. Nevertheless, static solutions also play a pivotal role in dissipative gravitational collapse of stars as they can represent an initial static configuration or a final static configuration \cite{bonnor,sharma1,sharma2}.

It is interesting to note that by relaxing the condition of a perfect fluid and allowing for pressure anisotropy and charge within the interior of the stellar distribution gives rise to observable and measurable properties of the star. Pressure anisotropy leads to arbitrarily large surface red-shifts   \cite{bhar1,bhar2,singh1} while the inclusion of charge results in the modification of the Buchdahl limit \cite{andy}. The linear equation of state $p=\alpha \rho$ has been generalized from observations in theoretical particle physics. There has been a wide spectrum of exact solutions of the field equations incorporating the so-called MIT bag model in which the equation of state is of the form $p=\alpha \rho-B$ with $B$ being the bag constant \cite{sunil2,sunil3,thiruk}. These solutions successfully predicted the observed masses and radii of compact objects with densities of the order of $10^{14} g~cm^{-3}$. With an ever growing interest in dark energy and its successful use in cosmological models, astrophysicists have now extended the range of $\alpha$ in $p = \alpha \rho$ to include $-1< \alpha < -1/3$. This regime incorporates the so-called dark stars \cite{bhar10,lobo}. Other exotic forms of matter which have appeared in the literature include the Chaplygin gas, Bose-Einstein condensates and the Hagedorn fluid \cite{far1,far2,bhar3,harko1,harko2}.

\begin{figure}[t] 
\begin{minipage}{\columnwidth}
\centering
\includegraphics[scale=0.5]{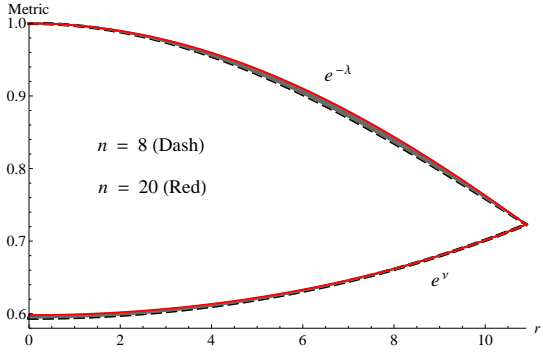}
\end{minipage}
\caption{Variation of metric potentials with radial coordinate $r$ for XTE J1739-285 $\big(n = 8-20,~M = 1.51M_\odot,~R=10.9km\big)$.}
\label{met}
\end{figure}

The notion of the four fundamental interactions being a manifestation of a single force has always attracted the interest of researchers in both fundamental particle physics and relativity. Higher dimensional theories of gravity have produced rich results so far as cosmic censorship is concerned  \cite{dad1,pankaj1,dadh}. Recently, there has been a surge in exact models of stars in Einstein-Gauss-Bonnet gravity, braneworld gravity as well as Lovelock gravity \cite{brian1,brian2,sudan1,megann1}. The connection between five-dimensional Kaluza-Klein geometries and electromagnetism has been widely studied. Embedding of four-dimensional space-times into higher dimensions is an invaluable tool in generating both cosmological and astrophysical models. 

In this article we are presenting a four dimensional spacetime embedded in five dimensional Pseudo Euclidean space known as class I. Karmarkar proposed a theory that any solutions of Einstein field equations that satisfies (\ref{con}) is said to be class I. For a neutral isotropic fluid sphere the solutions yield from (\ref{con}) is either Schwarzschild interior (1916) or Kholar-Chao (1965). However, it is now well known that inclusion of electric charge or pressure anisotropy or both lead to completely new class of solutions \cite{newton1,newton2,newton3,jiten,smit}. Therefore, in this paper we utilize the Karmarkar \cite{kar} condition which is a necessary and sufficient condition for a spherically symmetric line element to be of class I to generate exact solutions. In particular, our model incorporates anisotropic pressure to generate a family of new solutions.

\begin{figure}[t] 
\begin{minipage}{\columnwidth}
\centering
\includegraphics[scale=0.5]{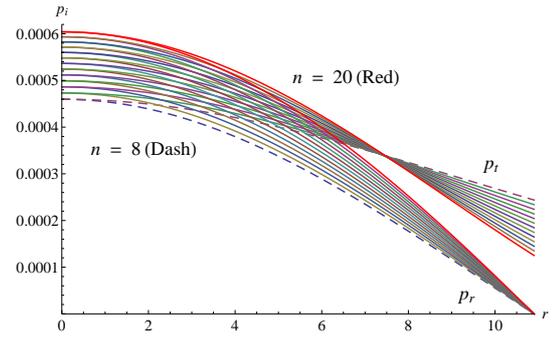}
\end{minipage}
\caption{Variation of interior pressures (km$^{-2}$) with radial coordinate $r$ for XTE J1739-217.}
\label{p}
\end{figure}

\begin{figure}[t] 
\begin{minipage}{\columnwidth}
\centering
\includegraphics[scale=0.5]{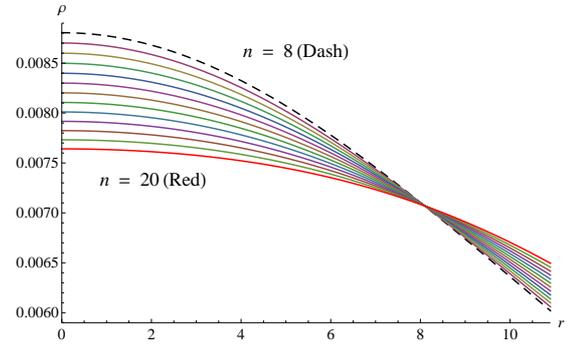}
\end{minipage}
\caption{Variation of density (km$^{-2}$)  with radial coordinate $r$ for XTE J1739-285.}
\label{rho}
\end{figure}

\section{Conditions for well-behaved solutions}

For well-behaved nature of the solutions for an anisotropic fluid sphere following conditions should be satisfied:

\begin{enumerate}
\item  The solution should be free from physical and geometric singularities, i.e. it should yield finite and positive values of the central pressure, central density and nonzero positive value of  $e^\nu|_{r=0}$ and  $e^\lambda|_{r=0}=1$.\\

\item  The causality condition should be obeyed i.e. velocity of sound should be less than that of light throughout the model. In addition to the above the velocity of sound should be decreasing towards the surface i.e.${d \over dr}~{dp_r \over d\rho}<0$  or   ${d^2 p_r\over d\rho^2}>0$   and ${d\over dr}{dp_t \over d\rho}<0$  or   ${d^2p_t \over d\rho^2}>0$  for $0\leq r\leq r_b$ i.e. the velocity of sound is increasing with the increase of density and it should be decreasing outwards.\\

\item  	The adiabatic index,  $\Gamma= {\rho+p_r \over p_r}~{dp_r \over d\rho}$  for realistic matter should be  $\Gamma>4/3$ for positive anisotropy.\\

\item   The anisotropy factor $\Delta$ should be zero at the center and increasing towards the surface.\\

\item   For a stable anisotropic compact star, $-1\leq v_t^2-v_r^2\leq 0$ must be satisfed \cite{herrera97}.\\

\end{enumerate}

\begin{figure} 
\begin{minipage}{\columnwidth}
\centering
\includegraphics[scale=0.5]{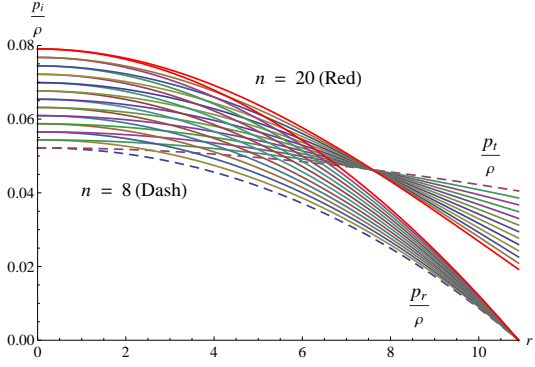}
\end{minipage}
\caption{Variation of pressure to density ratios  with radial coordinate $r$ for XTE J1739-285.}
\label{prho}
\end{figure}

\section{Einstein field equations of anisotropic fluid distributions}

The interior of the super-dense star is assumed to be described by the line element
\begin{equation}
ds^2 = e^{\nu(r)}dt^2 - e^{\lambda(r)}dr^2 - r^2(d\theta^2 + \sin^2{\theta}d\phi^2) \label{metric}
\end{equation}

The Einstein field equations for anisotropic fluid distribution are given as (in the unit $G=c=1$)
\begin{equation}
R^\mu_\xi-{1\over 2}R~g^\mu_\xi=-8\pi T^\mu_\xi
\end{equation}
where 
\begin{eqnarray}
T^\mu_\xi & = & (p_t+\rho)v^\mu v_\xi-p_t g^\mu_\xi+(p_r-p_t)\chi_\xi \chi^\mu
\end{eqnarray}
where  $R^\mu_\xi$  is Ricci tensor, $T^\mu_\xi$ is energy-momentum tensor, $R$ the scalar curvature, $p_r$ and $p_t$ denote radial and transverse pressures respectively, $\rho$ the density distribution , $v^\mu$   the four velocity and $\chi^\mu$  is the unit space-like vector in the radial direction.

Here $T^\mu_\xi$ is defined as
\begin{eqnarray}
T^\mu_\xi = \mbox{diag} \Big(\rho,~-p_r,~-p_t,~-p_t\Big)
\end{eqnarray}

\begin{figure}[t] 
\begin{minipage}{\columnwidth}
\centering
\includegraphics[scale=0.5]{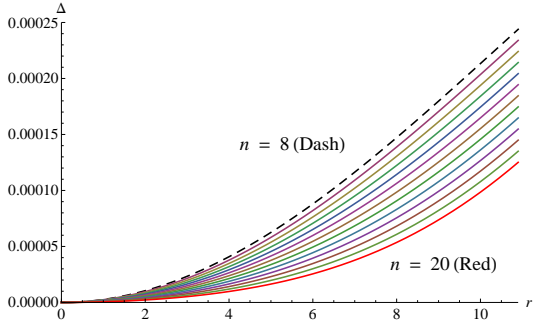}
\end{minipage}
\caption{Variation of anisotropy (km$^{-2}$) with radial coordinate $r$ for XTE J1739-285.}
\label{aniso}
\end{figure}

\begin{figure} 
\begin{minipage}{\columnwidth}
\centering
\includegraphics[scale=0.5]{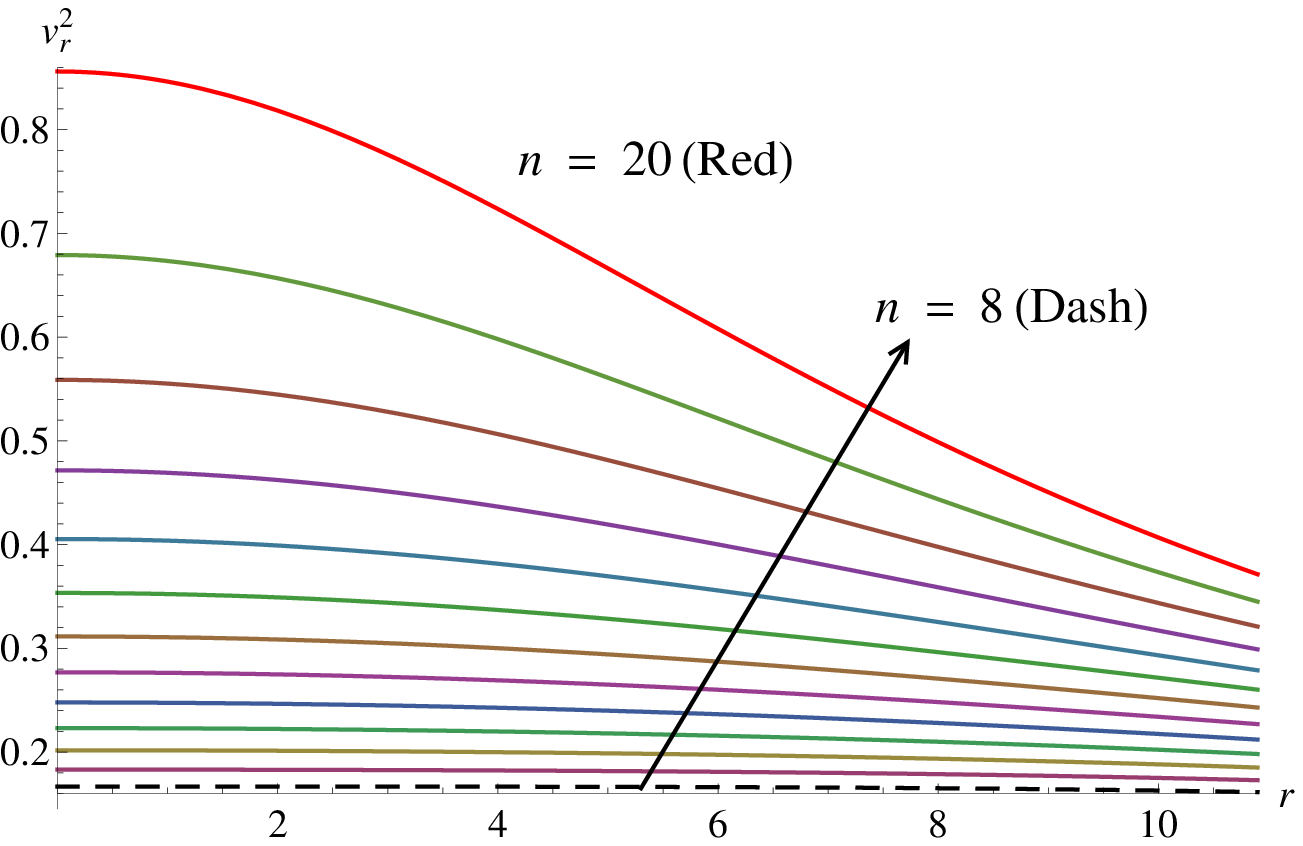}
\end{minipage}
\caption{Variation of $v_r^2$ with radial coordinate $r$ for XTE J1739-285.}
\label{sound}
\end{figure}

The Einstein field equations for the line element (\ref{metric}) are
\begin{eqnarray}\label{g3}
8\pi \rho &=&
\frac{1 - e^{-\lambda}}{r^2} + \frac{\lambda'e^{-\lambda}}{r} \label{g3a} \\ \nonumber \\
8\pi p_r &=&  \frac{\nu' e^{-\lambda}}{r} - \frac{1 - e^{-\lambda}}{r^2} \label{g3b}
\end{eqnarray}
\begin{eqnarray}
8\pi p_t &=& \frac{e^{-\lambda}}{4}\left(2\nu'' + {\nu'}^2  - \nu'\lambda' + \frac{2\nu'}{r}-\frac{2\lambda'}{r}\right) \label{g3c}
\end{eqnarray}

where primes represent differentiation with respect to the radial coordinate $r$. In generating the above field equations we have utilized geometrized units where the coupling constant and the speed of light are taken to be unity. Using Eqs. (\ref{g3b}) and (\ref{g3c}) we get

\begin{eqnarray}
\Delta &=& 8\pi (p_t-p_r) \nonumber \\
&=& e^{-\lambda}\left[{\nu'' \over 2}-{\lambda' \nu' \over 4}+{\nu'^2 \over 4}-{\nu'+\lambda' \over 2r}+{e^\lambda-1 \over r^2}\right] \label{del}
\end{eqnarray}

If the metric given in (\ref{metric}) satisfies the \cite{kar} condition , it can represent an embedding class I spacetime i.e. 
\begin{equation}
R_{1414}={R_{1212}R_{3434}+R_{1224}R_{1334} \over R_{2323}}\label{con}
\end{equation}
with $R_{2323}\neq 0$, \cite{pandey}. This condition leads to a differential equation given by
\begin{equation}
{2\nu'' \over \nu'}+\nu'={\lambda' e^\lambda \over e^\lambda-1}\label{dif1}
\end{equation}

\begin{figure} 
\begin{minipage}{\columnwidth}
\centering
\includegraphics[scale=0.5]{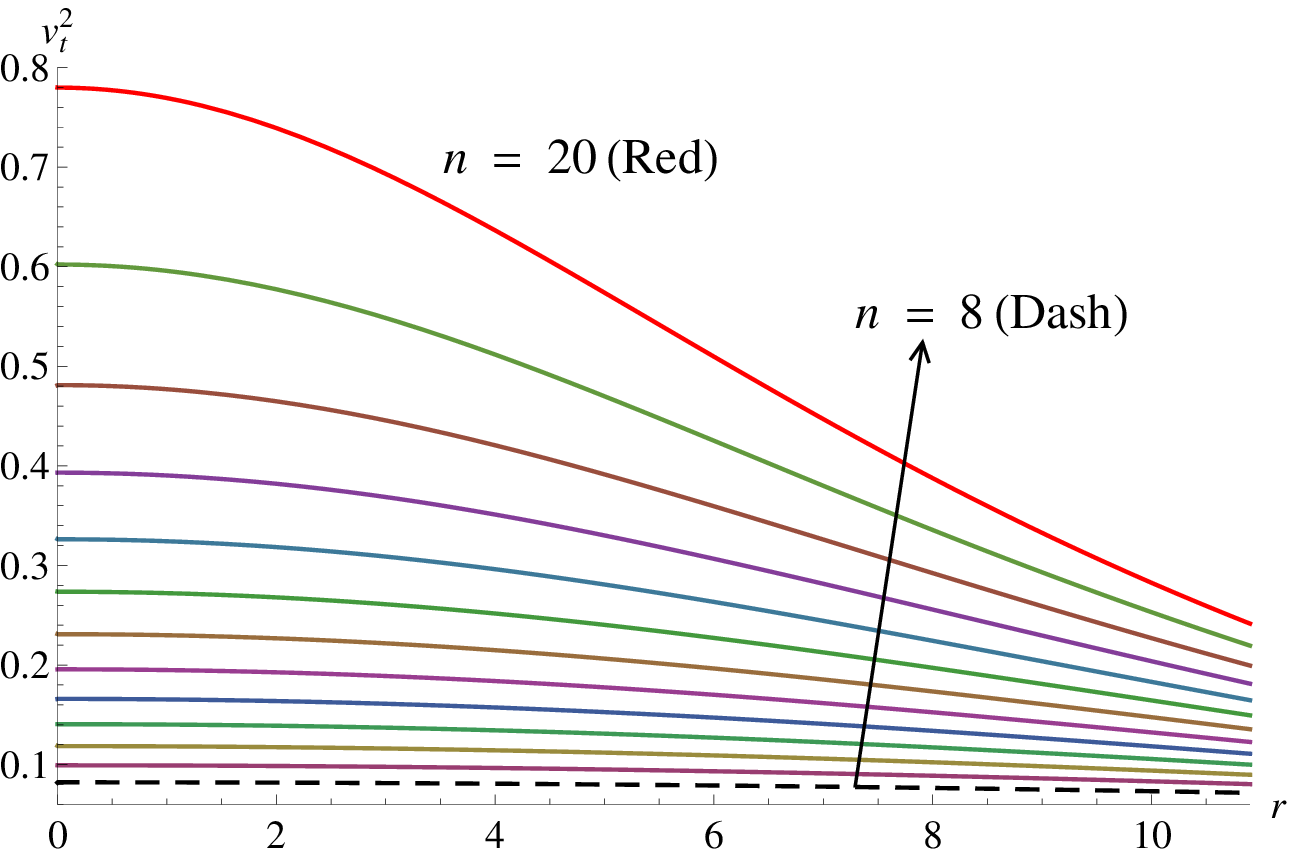}
\end{minipage}
\caption{Variation of $v_t^2$ (km$^{-2}$)  with radial coordinate $r$ for XTE J1739-285.}
\label{vt}
\end{figure}

\begin{figure} 
\begin{minipage}{\columnwidth}
\centering
\includegraphics[scale=0.5]{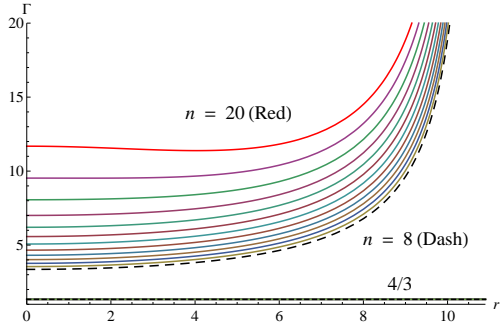}
\end{minipage}
\caption{Variation of relativistic adiabatic index  with radial coordinate $r$ for XTE J1739-285.}
\label{gamma}
\end{figure}

\begin{figure} 
\begin{minipage}{\columnwidth}
\centering
\includegraphics[scale=0.5]{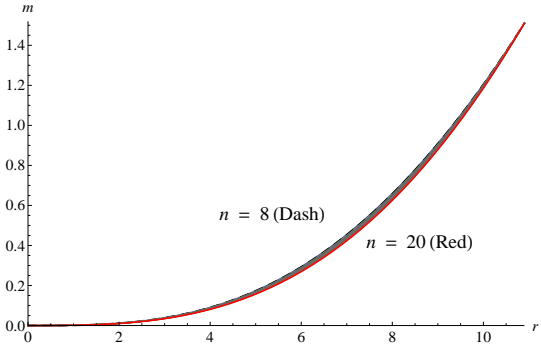}
\end{minipage}
\caption{Variation of interior mass with radial coordinate $r$ for XTE J1739-285.}
\label{mas}
\end{figure}

\begin{figure} 
\begin{minipage}{\columnwidth}
\centering
\includegraphics[scale=0.5]{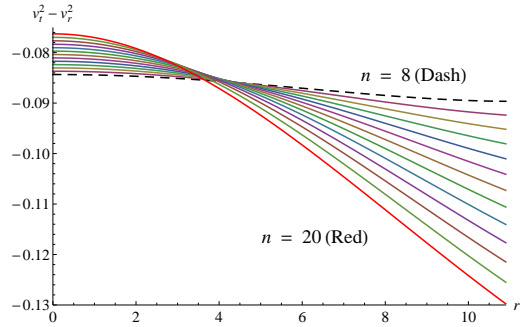}
\end{minipage}
\caption{Variation of stability factor ~$v_t^2-v_r^2$ ~ with radial coordinate $r$ for XTE J1739-285.}
\label{stab}
\end{figure}

\begin{figure} 
\begin{minipage}{\columnwidth}
\centering
\includegraphics[scale=0.5]{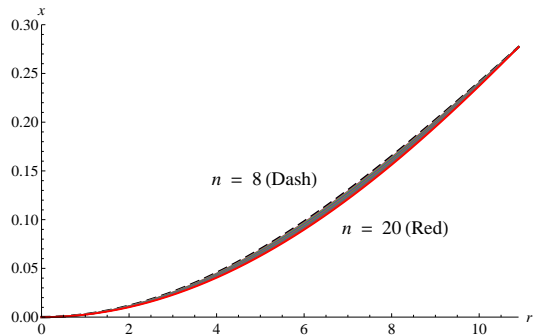}
\end{minipage}
\caption{Variation of compactness parameter with radial coordinate $r$ for XTE J1739-285.}
\label{com}
\end{figure}

\begin{figure} 
\begin{minipage}{\columnwidth}
\centering
\includegraphics[scale=0.5]{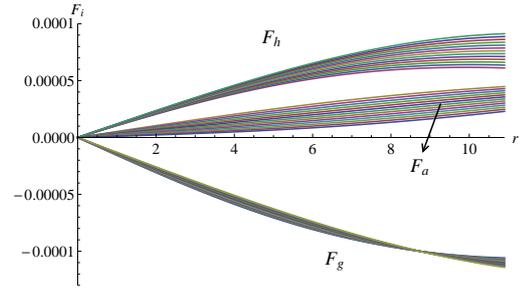}
\end{minipage}
\caption{Balancing of different forces in TOV equation for a static configuration for XTE J1739-285 are plotted with radial coordinate $r$.}
\label{tov}
\end{figure}

\begin{figure} 
\begin{minipage}{\columnwidth}
\centering
\includegraphics[scale=0.5]{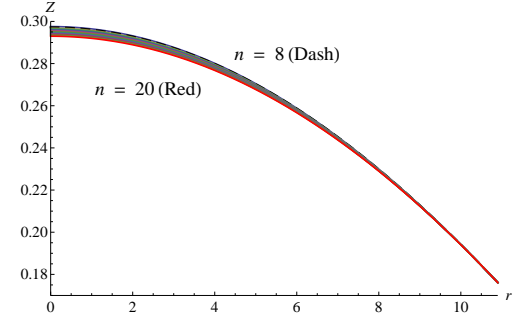}
\end{minipage}
\caption{Variation of red-shift with radial coordinate $r$ for XTE J1739-285.}
\label{red}
\end{figure}

\begin{figure} 
\begin{minipage}{\columnwidth}
\centering
\includegraphics[scale=0.5]{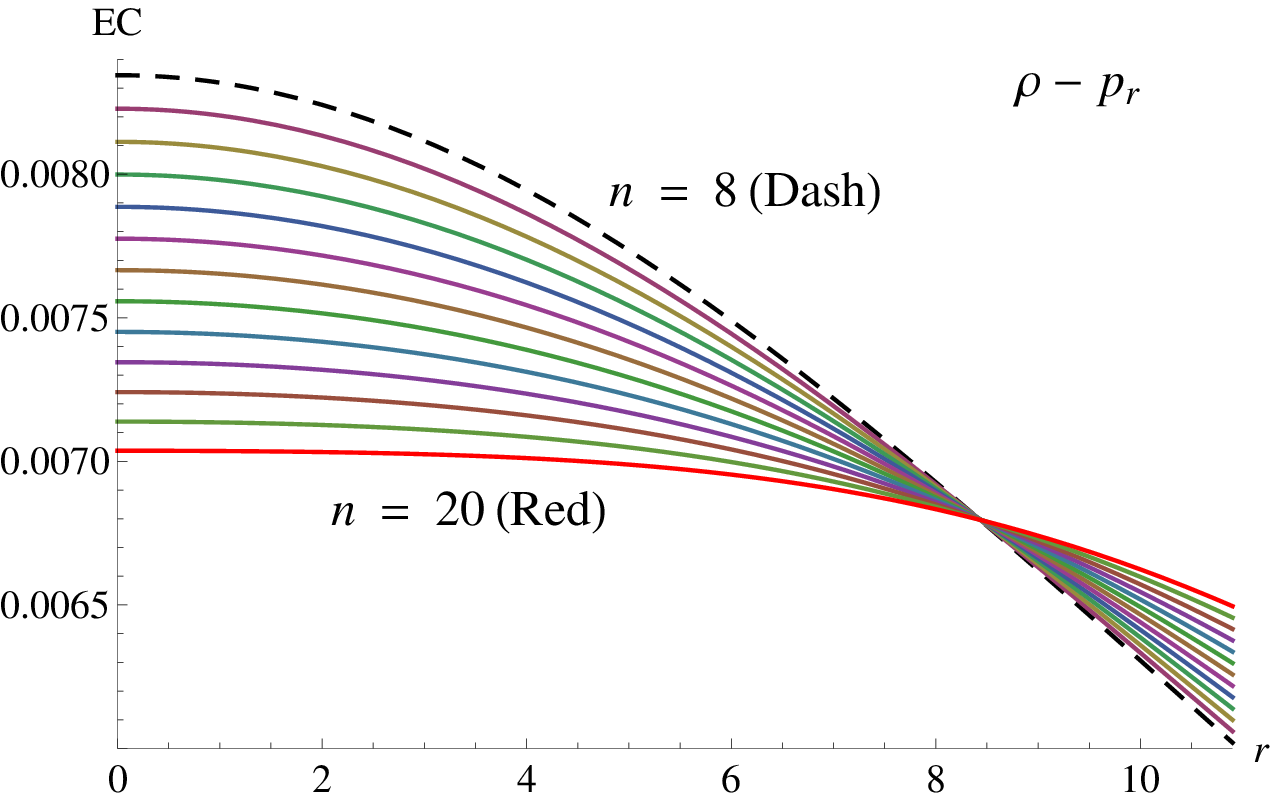}
\end{minipage}
\caption{Variation of $\rho-p_r$ (km$^{-2}$) with radial coordinate $r$ for XTE J1739-285.}
\label{ec1}
\end{figure}

\begin{figure} 
\begin{minipage}{\columnwidth}
\centering
\includegraphics[scale=0.5]{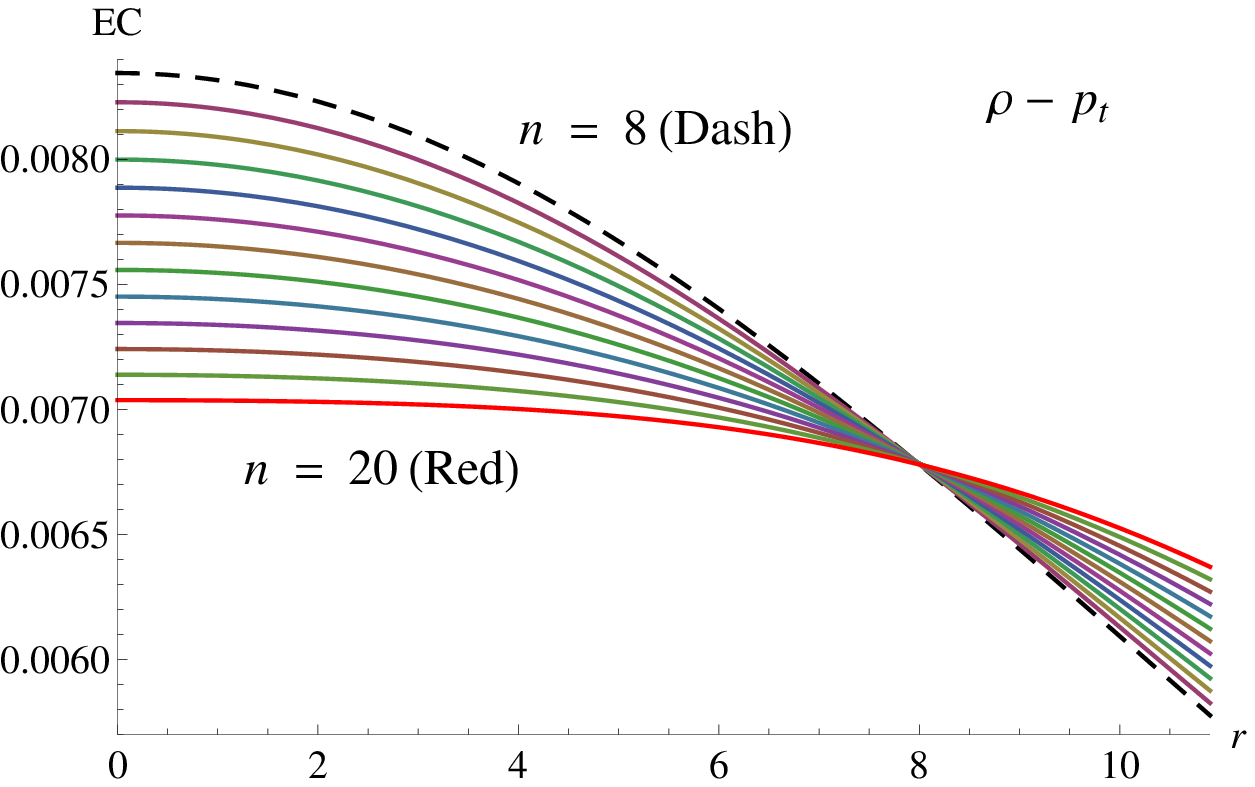}
\end{minipage}
\caption{Variation of $\rho-p_t$ (km$^{-2}$) with radial coordinate $r$ for XTE J1739-285.}
\label{ec2}
\end{figure}

\begin{figure} 
\begin{minipage}{\columnwidth}
\centering
\includegraphics[scale=0.5]{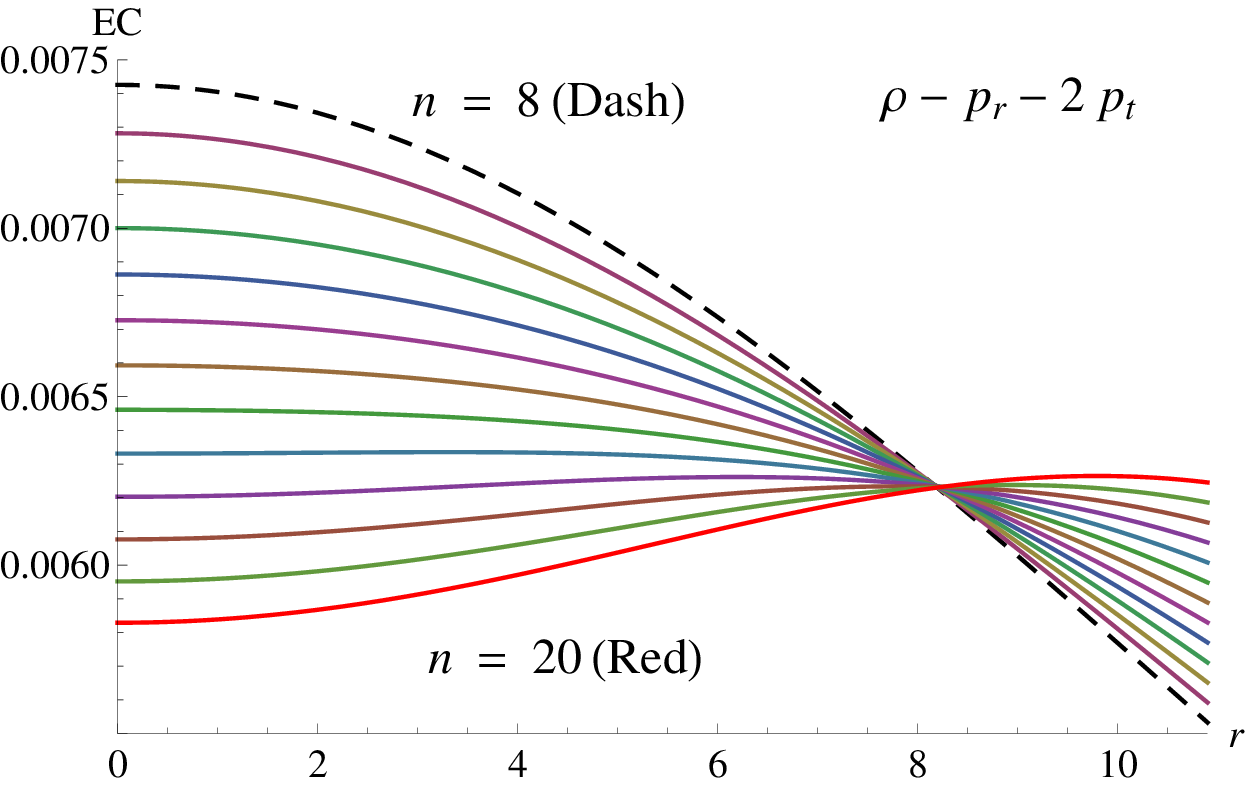}
\end{minipage}
\caption{Variation of $\rho-p_r-2p_t$ (km$^{-2}$) with radial coordinate $r$ for XTE J1739-285.}
\label{ec3}
\end{figure}

On integration we get the relationship between $\nu$ and $\lambda$ as
\begin{equation}
e^{\nu}=\left(A+B\int \sqrt{e^{\lambda}-1}~dr\right)^2\label{nu1}
\end{equation}
where $A$ and $B$ are constants of integration.

By using (\ref{nu1}) we can rewrite (\ref{del}) as
\begin{eqnarray}
\Delta = {\nu' \over 4e^\lambda}\left[{2\over r}-{\lambda' \over e^\lambda-1}\right]~\left[{\nu' e^\nu \over 2rB^2}-1\right] \label{del1}
\end{eqnarray}
Here $\Delta=8\pi (p_t-p_r)$ is the measure of anisotropy.

\section{Generating a new family of embedding class I solution}

To solve the above equation (\ref{nu1}), we assumed a new metric potential $g_{rr}$ given by
\begin{equation}
e^\lambda =1+ar^2 (1+br^2)^n \label{elam}
\end{equation}
where  $a$ and $b$ are constants with dimension of length$^{-2}$.

Using the metric potential (\ref{elam}) in (\ref{nu1}), we get
\begin{eqnarray}
e^\nu = \left[A+\frac{\sqrt{a} B (b r^2+1)^{n/2+1}}{b (n+2)}\right]^2 ~~~\mbox{where}~~~n \ne -2\label{enu}
\end{eqnarray}

Using (\ref{elam}) and (\ref{enu}), we can rewrite the expression of density, $p_r$, $\Delta$ and $p_t$ as
\begin{eqnarray}
8\pi \rho & = & \frac{a (b r^2+1)^{n-1}}{\big[a r^2(b r^2+1)^n+1\big]^2} \times \Big[a r^2 (b r^2+1)^n \nonumber\\
& & +b r^2 \big\{a r^2 (b r^2+1)^n+2 n+3\big\}+3\Big]\\
8\pi p_r & = & \frac{\sqrt{a(b r^2+1)^n}~[a r^2(b r^2+1)^n+1]^{-1}}{B (b r^2+1) \sqrt{a (b r^2+1)^n}+A b (n+2)}\times \nonumber\\
&& \bigg[b \Big\{B \left(-a r^2(b r^2+1)^n+2 n+4\right)-A (n+2) \nonumber \\
&&  \sqrt{a(b r^2+1)^n}\Big\}-a B(b r^2+1)^n\bigg]\\
\Delta &=& {r^2 \Big[a b r^2(b r^2+1)^n+a(b r^2+1)^n-b n\Big] \over (b r^2+1) \big[(a r^2(b r^2+1)^n+1\big]^2} \times \nonumber \\
& & \bigg[B (b r^2+1) \sqrt{a(b r^2+1)^n}+A b (n+2) r\bigg]^{-1} \nonumber \\
&& \bigg[a(b r^2+1)^n \Big\{B (b r^2+1) \sqrt{a(b r^2+1)^n} \nonumber \\
&& +A b (n+2)\Big\}-b B (n+2) \sqrt{a(b r^2+1)^n}\bigg]\\
8\pi p_t &=& 8\pi p_r+\Delta
\end{eqnarray}

Now the pressure and density gradients can be written as
\begin{eqnarray}
8\pi {d\rho \over dr} & = & -\frac{2 a r [f_1(r)+f_2(r)] (b r^2+1)^{n-2}}{[a r^2(b r^2+1)^n+1]^3}\\
8\pi {dp_r \over dr} & = & {2 \sqrt{a r^2(b r^2+1)^n}  \over (b r^2+1) \{a r^2(b r^2+1)^n+1\}^2 } \times \nonumber \\
&& \bigg[B(b r^2+1) \sqrt{a(b r^2+1)^n}+A b (n+2)\bigg]^{-2} \nonumber \\
&& \bigg[B^2f_5(r) (b r^2+1) \sqrt{a r^2 (b r^2+1)^n}+f_3(r) \nonumber \\
&& +A b B f_4(r) (n+2) r\bigg] \\
8\pi {dp_t \over dr} & = & \frac{1}{\beta(r)^2 (b r^2+1)^2 \big[a r^2 (b r^2+1)^n+1\big]^3} \times \nonumber \\
&& \bigg[ -2 bf_6(r)~ r [a r^2 (b r^2+1)^n+1] \times \nonumber \\
&& \Big\{B (b r^2+1) \sqrt{a r^2 (b r^2+1)^n}+A b (n+2) r\Big\} \nonumber \\
&& -\frac{\alpha(r)f_6(r) (b r^2+1) \big\{a r^2 (b r^2+1)^n+1\big\}}{\sqrt{a (b r^2+1)^n}} -\nonumber \\
&& 4 a\beta(r) f_6(r) r(b r^2+1)^n \big\{b (n+1) r^2+1\big\}+ \nonumber \\
&& (b r^2+1) \big\{a r^2 (b r^2+1)^n+1\big\} \beta(r) \bigg\{f_7(r) \nonumber \\
&& -\frac{f_{9}(r) \sqrt{a r^2 (b r^2+1)^n}}{r^2}-f_{8}(r)\bigg\}\bigg]
\end{eqnarray}
where
\begin{eqnarray}
f_1(r) & = & a (b r^2+1)^n \big\{a r^2 (b r^2+1)^n+5\big\}+ \nonumber \\
&& \big[2 a r^2 (b r^2+1)^n \big\{a r^2 (b r^2+1)^n+5\big\} \nonumber \\
&& +n \big\{(3 a r^2 (b r^2+1)^n-5\big\} \big] \\
f_2(r) & = & b^2r^2\Big[2 n^2 \big\{(a r^2(b r^2+1)^n-1\big\}+\\
&& a r^2 (b r^2+1)^n \big\{a r^2 (b r^2+1)^n+5\big\}+\nonumber \\
&& n \big\{(5 a r^2 (b r^2+1)^n-3\big\} \Big]\\
f_3(r) & = & A^2 b^2 (n+2)^2 \sqrt{a r^2 (b r^2+1)^n} \Big\{a b r^2 (b r^2+1)^n \nonumber \\
&& +a (b r^2+1)^n-b n\Big\} \\
f_4(r) & = & 4 a b(b r^2+1)^n \big\{a r^2 (b r^2+1)^n-n-1\big\}+ \nonumber \\
&& 2 a^2 (b r^2+1)^{2 n}+b^2\big[n \{2-6 a r^2 (b r^2+1)^n\} \nonumber \\
&& +n^2 \big\{1-a r^2 (b r^2+1)^n\big\}+ 2 a r^2 (b r^2+1)^n \nonumber \\
&& \big\{a r^2 (b r^2+1)^n-2\big\}\big]\\
f_5(r) & = & a b (b r^2+1)^n \big\{2 a r^2 (b r^2+1)^n-3 n-4\big\} + \nonumber \\
&&  a^2 (b r^2+1)^{2 n}-b^2 \Big[8 a r^2 (b r^2+1)^n+4 \nonumber \\
&& +2 a n^2 r^2 (b r^2+1)^n-a^2 r^4 (b r^2+1)^{2 n}+\nonumber \\
&& n \big\{2 + 9 a r^2 (1 + b r^2)\big\}^n\Big]  \\
f_6(r) & = & b B (n+2) (b (n+2) r^2+2) \sqrt{a r^2 (b r^2+1)^n} \nonumber \\
&& -a (b r^2+1)^n \Big[B (1-b^2 r^4) \sqrt{a r^2 (b r^2+1)^n}\nonumber \\
&& +A b (n+2) r (b (n+1) r^2+1)\Big] \\
f_7(r) & = & 2 b^2 B (n+2)^2 r \sqrt{a r^2 (b r^2+1)^n}+\frac{b B (n+2)}{b r^3+r} \nonumber \\
&& \big\{b (n+1) r^2+1\big\} \big\{b (n+2) r^2+2\big\}\nonumber \\
&& \sqrt{a r^2 (b r^2+1)^n} \times\\
f_8(r) &=& 2 a b n r (b r^2+)^{n-1} \Big[A b (n+2) r \big\{b (n+1) r^2+1\big\} \nonumber \\
&& +B (1-b^2 r^4) \sqrt{a r^2 (b r^2+1)^n}\Big] \\
f_9(r) &=& A b (n+2) \big\{3 b (n+1) r^2+1\big\} \sqrt{a r^2 (b r^2+1)^n} \nonumber \\
&& -a B r (b r^2+1)^n \big\{b^2 (n+5) r^4-b n r^2-1\big\} \\
\beta(r) &=& B (b r^2+1) \sqrt{a r^2 (b r^2+1)^n}+A b (n+2) r  \\
\alpha(r) &=& A b (n+2) \sqrt{a (b r^2+1)^n}+a B \big\{b (n+3) r^2+1\big\} \nonumber\\
&& \times (b r^2+1)^n \\
\end{eqnarray}

\begin{figure} 
\begin{minipage}{\columnwidth}
\centering
\includegraphics[scale=0.25]{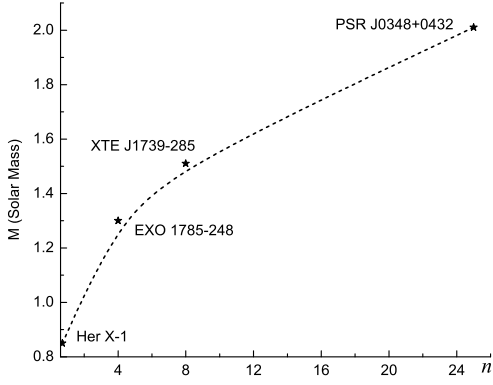}
\end{minipage}
\caption{Variation of mass with the star generating index $n$ plotted from Her X-1, EXO 1785-248, XTE J1739-285 and PSR J0348+0432.}
\label{star}
\end{figure}

\section{Properties of the new family solution}

The central values of pressures and density are given by
\begin{eqnarray}
8\pi p_r(r=0) & = & 8\pi p_t(r=0) \nonumber \\
&  = & \frac{\sqrt{a}}{\sqrt{a} B+A b (n+2)}\bigg[b \big\{B (2 n+4) \nonumber \\
&& -\sqrt{a} A (n+2)\big\}-a B\bigg]>0\label{pc}\\
\rho(r=0) & = & 3a>0;~~\forall~a>0
\end{eqnarray}

To satisfy Zeldovich's condition at the interior, $p_r/\rho$ at center must be $\le 1$. Therefore,
\begin{eqnarray}
\frac{b \big\{B (2 n+4)-\sqrt{a} A (n+2)\big\}-a B}{3\sqrt{a}\big[\sqrt{a} B+A b (n+2)\big]} \le 1 \label{zel}
\end{eqnarray}

On using (\ref{pc}) and (\ref{zel}) we get a constraint on $B/A$ given as
\begin{equation}
\frac{\sqrt{a}~ b (n+2)}{2 b n+4 b-a}<{B \over A} \le \frac{4\sqrt{a}~ b (n+2)}{2 b n+4 b-4a}
\end{equation}

Now the velocity of sound inside the stellar interior can be determined by using
\begin{equation}
v_r^2={dp_r/dr \over d\rho/dr},~~~v_t ^2={dp_t/dr \over d\rho/dr}
\end{equation}

The relativistic adiabatic index for an anisotropic fluid sphere is given by
\begin{equation}
\Gamma = {\rho+p_r \over p_r}~{dp_r \over d\rho}
\end{equation}

For a static configuration at equilibrium $\Gamma$ has to be more than $4/3$.

The modified Tolman-Oppenheimer-Volkoff (TOV) equation for anisotropic fluid distribution was given by \cite{ponce} as
\begin{eqnarray}
-{M_g(\rho+p_r) \over r^2}~e^{(\lambda-\nu)/2}-{dp_r \over dr}+{2\Delta \over r}=0 \label{tove}
\end{eqnarray}

where 
\begin{eqnarray}
M_g(r) &=& {1\over 2}~r^2 \nu' e^{(\nu-\lambda)/2}
\end{eqnarray}

The above equation (\ref{tove}) can be written in terms of balanced force equation due to anisotropy ($F_a$), gravity ($F_g$) and hydrostatic ($F_h$) i.e.
\begin{equation}
F_g+F_h+F_a=0 \label{forc}
\end{equation}

Here
\begin{eqnarray}
F_g & = & -{M_g(\rho+p_r) \over r^2}~e^{(\lambda-\nu)/2}\\
F_h &=& -{dp_r \over dr}\\
F_a &=& {2\Delta \over r}\\
\end{eqnarray}
The TOV equation (\ref{forc}) can be represented by the figure showing that the forces are counter balanced to each other  Fig. (\ref{tov}).

\section{Matching of physical boundary conditions}

Assuming the exterior spacetime to be the Schwarzschild exterior solution which has to match smoothly with our interior solution and is given by
\begin{eqnarray}
ds^2 &=& \left(1-{2M\over r}\right) dt^2-\left(1-{2M\over r}\right)^{-1}dr^2 \nonumber\\
& & -r^2(d\theta^2+\sin^2 \theta d\phi^2) \label{ext}
\end{eqnarray}

\begin{table*}
\caption{Parameters of four well-known compact stars that gives mass and radius compatible to observed values and corresponds to well behaved solution.}
\label{tab}
\begin{tabular*}{\textwidth}{@{\extracolsep{\fill}}lrrrrrrrrl@{}}
\hline
\multicolumn{1}{c}{$n$} & \multicolumn{1}{c}{$a$ (km$^{-2}$)} & \multicolumn{1}{c}{$b$ (km$^{-2}$)} & \multicolumn{1}{c}{$A$} & \multicolumn{1}{c}{$B$ (km$^{-1}$)} & \multicolumn{1}{c}{$r_b$ (km)} & \multicolumn{1}{c}{$M/M_\odot$} & \multicolumn{1}{c}{$x(r_b)=2M/r_b$} & \multicolumn{1}{c}{$Z_s$} & Object \\
\hline
0.7 & 0.0038724 & 0.001 & 0.178738 & 0.0282792 & 8.1 & 0.85 & 0.209877 & 0.125 & Her X-1 \\
4 & 0.00416517 & 0.0008 & 0.346286 & 0.0306226 & 8.85 & 1.3 & 0.293785 & 0.189958 & EXO 1785-248  \\
8 & 0.00293489 & 0.0001 & $-$0.537384 & 0.0241454 & 10.9 & 1.51 & 0.277064049 & 0.176115812 & XTE J1739-285 \\
25 & 0.00409351 & 0.00005 & $-$0.61619 & 0.0274786 & 11 & 2.01 &  0.365454402 & 0.255360854 & PSR J0348+0432\\
\hline
\end{tabular*}
\end{table*}

By matching the first and second fundamental forms the interior solution (\ref{metric}) and exterior solution (\ref{ext}) at the boundary $r=r_b$ (Darmois-Isreali condition) we get
\begin{eqnarray}
e^{\nu_b} &=& 1-{2M \over r_b} = \left[A+\frac{\sqrt{a} B (b r_b^2+1)^{n/2+1}}{b (n+2)}\right]^2\label{bou1}\\
e^{-\lambda_b} &=& 1-{2M \over r_b} = \Big[1+ar_b^2(1+br_b^2)^n\Big]^{-1} \label{bou2}\\
p_r(r_b) &=& 0 \label{bou3}
\end{eqnarray}

Using the boundary condition (\ref{bou1}-\ref{bou3}),  we get
\begin{eqnarray}
B &=& b A(n+2) \sqrt{a (b r_b^2+1)^n} \times \nonumber \\
&& \bigg[b\Big\{-a r_b^2 (b r_b^2+1)^n+2 n+4\Big\}-a (b r_b^2+1)^n\bigg] \label{b}\\
A & = & \sqrt{a r_b^2 (b r_b^2+1)^n+1} \times \nonumber \\
&& \bigg[1+a (b r_b^2+1)^{n+1} ~\bigg\{ b \Big[-a r_b^2 (b r_b^2+1)^n +2 n+4 \Big] \nonumber \\
 && -a (b r_b^2+1)^n\bigg\} \bigg]\\
a &=& \frac{1}{r_b^2 (b r_b^2+1)^n} \left(\frac{1}{1-2M/r_b}-1 \right) \label{a}
\end{eqnarray}
and we have chosen $b,~n,~M$ and $r_b$ as free parameters and the rest of the constants $a,~A$ and $B$ are determined from the Eqs. (\ref{b}-\ref{a}).

Now the gravitational red-shift of the stellar system is given by
\begin{eqnarray}
Z(r) & = & \left[A+\frac{\sqrt{a} B (b r^2+1)^{n/2+1}}{b (n+2)}\right]^{-1}-1
\end{eqnarray}

The mass-radius relation and compactness parameter of the solution can be determined using the equation given below:
\begin{eqnarray}
m(r) & = & 4\pi \int_0^r \rho r^2 dr = \frac{a r^3 (b r^2+1)^n}{2 a r^2 (b r^2+1)^n+2}\\
x(r) &=& {2m(r) \over r} = \frac{2a r^2 (b r^2+1)^n}{2 a r^2 (b r^2+1)^n+2}
\end{eqnarray}

\section{Results and conclusions}

It has been observed that the physical parameters $\big(e^{-\lambda},~p_r,$ $p_t,~\rho,$ $~p_r/\rho,~p_t/\rho,~v_r^2,~v_t^2,~Z\big)$ are free from central singularities and monotonically decreasing outward (Figs. \ref{met}, \ref{p}, \ref{rho}, \ref{prho}, \ref{sound}, \ref{vt}, \ref{red}). However $e^\nu$, anisotropy and $\Gamma$ are increasing outward and also well-behaved (Figs. \ref{met}, \ref{aniso}, \ref{gamma}). 

Furthermore, our presented solution satisfies all the energy condition which are needed by a physically possible configurations. The Null Energy Condition $\big(\rho-p_i \ge 0\big)$, Dominant Energy Condition $\big(\rho-p_i \ge 0$, $\rho \ge 0\big)$ and Strong Energy Condition $\big(\rho-p_i \ge 0$, $\rho-p_r-2p_t \ge 0\big)$ is shown in Figs. \ref{ec1}, \ref{ec2} and \ref{ec3}. For a stable configuration, the stability factor $v_t^2-v_r^2$ must lies in between $-1$ to 0, which is again satisfied by our presented solutions (Fig. \ref{stab}). For a non-collapsing stellar configuration, the adiabatic index must also be more than 4/3 for positive values of anisotropy, Fig. \ref{gamma}. Furthermore, we can analyze all the forces acting on the physical system via Eq. (\ref{tove}) and we expect to counter-balanced for a static stellar configuration (Fig. \ref{tov}). The mass and the compactness parameter are also monotonically increase from the center to the surface of the star and the compactness parameter is also within the Buchdahl limit (Figs. \ref{mas}, \ref{com}).

For XTE J1739-285, the well behaved region of index $n$ ranges from 8 to 20. For $n<8$, the trend of sound speed increases and for $n>20$, the causality condition is violated. For PSR J0348+0432, the well-behaved region of $n$ is from 24 to 54, where $n<24$ gives increasing sound speed and $n>54$ yields violation of causality condition. For Her X-1, the well-behaved region belongs to $n=0.8$ to $n=2.7$, where $n<0.8$ and $n>2.7$ imply increasing sound speed and violation of causality condition respectively. Similarly, for EXO 1785-248 the well-behaved region is $n=1.5$ to $n=4$, where $n<1.5$ and $n>4$ imply increasing sound speed and violation of causality condition respectively. From the present analysis, we have observed that the equation of states are softer for small values of $n$ and vice versa. The adiabatic index for a particular star at the center increases with increase in $n$ i.e. the equation of state gets stiffer with the increase of $n$. For Her X-1, the adiabatic index at the center are 3.36 ($n=0.8$) and 5.47 ($n=2.7$); for EXO 1785-248 the adiabatic index at the center are 3.13 ($n=1.5$) and 10.38 ($n=4$); for XTE J1739-285 the adiabatic index at the center are 3.34 ($n=8$) and 11.65 ($n=20$); for PSR J0348+0432 the adiabatic index at the center are 2.7 ($n=24$) and 9.35 ($n=54$). However, for all the stars, for those values of $n$ which is less or more than the above mentioned limits, the energy conditions are indeed satisfied although not well-behaved. Since, for the different ranges of $n$ we may generate different stars with all degrees of suitability, the `$n$' is named as ``{\it star generating index}''. The dependence of mass on star generating index is shown in Fig. \ref{star}.  

For all the presented stars, with the star generating index $n$ within its well-behaved values, the central values of density, red-shift and energy conditions decrease, however, the central values of pressure, pressure to density ratio, speed of sound and adiabatic index increase.

The masses and radii of the chosen compact star candidates are matched with  \cite{abu} for Her X-1, \cite{ozel} for EXO 1785-248, \cite{zhang} for XTE J1739-285 and \cite{anto} for PSR J0348+0432. The parameters with masses and radii of each star are given in Table \ref{tab}. Further, for $n=-2$, we discovered the well-behaved solution \cite{newton1}. Furthermore, the present solution as- ymptotically approaches to \cite{mau} when $n \rightarrow \infty$ and suitable choice of parameter $b=2A/n$. Further, the analysis of the adiabatic index obtained of Maurya et al. \cite{mau} solution gives stiffer equation state however, in our solution the equation of state may be stiffer or softer depending upon the values of star generating index $n$.

\begin{acknowledgements}
Authors are grateful to the anonymous referee(s) for rigorous review, constructive comments and useful suggestions.
The authors also acknowledge their gratitude to Air Vice Marshal S.P. Wagle VM, the Deputy Commandant, NDA, for his motivation and encouragement.
\end{acknowledgements}

\end{document}